# Explainable AI Methods for Multi-Omics Analysis: A Survey

AHMAD HUSSEIN, School of Computer Science, Faculty of Engineering and Information Technology, University of Technology Sydney, Australia

MUKESH PRASAD, School of Computer Science, Faculty of Engineering and Information Technology, University of Technology Sydney, Australia

ALI BRAYTEE, School of Computer Science, Faculty of Engineering and Information Technology, University of Technology Sydney, Australia

Advancements in high-throughput technologies have led to a shift from traditional hypothesis-driven methodologies to data-driven approaches. Multi-omics refers to the integrative analysis of data derived from multiple 'omes', such as genomics, proteomics, transcriptomics, metabolomics, and microbiomics. This approach enables a comprehensive understanding of biological systems by capturing different layers of biological information. Deep learning methods are increasingly utilized to integrate multi-omics data, offering insights into molecular interactions and enhancing research into complex diseases. However, these models, with their numerous interconnected layers and nonlinear relationships, often function as black boxes, lacking transparency in decision-making processes. To overcome this challenge, explainable artificial intelligence (xAI) methods are crucial for creating transparent models that allow clinicians to interpret and work with complex data more effectively. This review explores how xAI can improve the interpretability of deep learning models in multi-omics research, highlighting its potential to provide clinicians with clear insights, thereby facilitating the effective application of such models in clinical settings.

## 1 Introduction

Omics is a broad field that involves the analysis of biological information interactions across different omes, such as genomics, proteomics, transcriptomics, metabolomics, and microbiomics [20]. Genomics is the most established among omics fields, it focuses on identifying genetic variations linked to disease, treatment response, or patient prognosis in medical research[20] . Transcriptomics involves studying gene transcription and transcriptional

Authors' Contact Information: Ahmad Hussein, Ahmad.Hussein@Student.uts.edu.au, School of Computer Science, Faculty of Engineering and Information Technology, University of Technology Sydney, Ultimo, NSW, Australia; Mukesh Prasad, mukesh.prasad@uts.edu.au, School of Computer Science, Faculty of Engineering and Information Technology, University of Technology Sydney, Ultimo, NSW, Australia; Ali Braytee, ali.braytee@uts.edu.au, School of Computer Science, Faculty of Engineering and Information Technology, University of Technology Sydney, Ultimo, NSW, Australia.

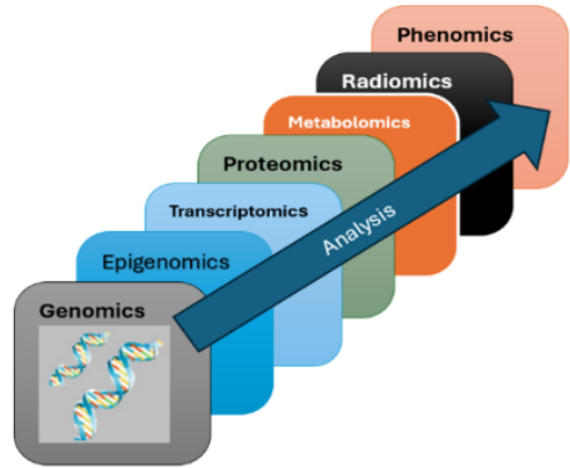

Fig. 1. A broad interpretation of multi-omics involves the comprehensive collection of omics data from various molecular levels, including the genome, transcriptome, proteome, epigenome, metabolome, microbiome, among others. This data is then subjected to integrated analysis to gain deeper insights into biological processes and molecular mechanisms [28]
.

regulation within cells. Proteomics focuses on analyzing the composition of cells, tissues, or biological proteins, as well as understanding their dynamic changes, with proteins as the primary research target. The term "microbiome" encompasses the genomes of microorganisms, including bacteria, archaea, lower or higher eukaryotes, and viruses, including their entire environment [28]. In addition to the progress in genomics, transcriptomics, and epigenomics integration approaches, radiomics— the extraction of various quantitative features from medical images— has proven to be a keyway to study cancer imaging phenotypes, reflecting underlying gene expression patterns. Therefore, integrating radiomics with other omics can provide meaningful knowledge in cancer diagnosis, prognosis, and treatment [60]. Combining data from multi-omics (Figure 1) such as genomics, transcriptomics, proteomics, and metabolomics is a researched systems biology approach aiming to comprehensively understand growth, adaptation, development, and disease progression, particularly with cancer, which is characterized by its heterogeneity. This approach is facilitated by high-throughput techniques, as well as the availability of large sample sets of multi-omics data, in addition to the development of advanced tools and methods for data integration and interpretation [46, 54].

Diseases can be caused by numerous simultaneous changes in cellular and molecular dynamics, including alterations in gene and protein expression, metabolic pathways, and tissue cell populations. An integrative approach to studying these complex changes and interactions can lead to a more comprehensive understanding of immunology, covering aspects such as viral replication inhibition, the generation of protective immune responses, pathogen evasion of innate and adaptive immunity, and differences in susceptibility among individuals and populations. Cancer stands as one of the primary causes of death globally, claiming millions of lives every year. Cancer biology serves as a crucial area of study aimed at understanding the mechanisms behind cancer development, progression, and treatment response. Researchers can now utilize the integration of multi-omics data to unravel the complex nature of cancer heterogeneity, tumorigenesis, and the discovery of anticancer drugs. This systematic approach involves thorough examination and analysis of cancer from diverse angles, such as genetics, epigenetics, signalling networks, cellular behaviour, clinical presentation, and epidemiology. For such integrative approaches, Deep Learning is assuming a particularly advantageous position in enhancing data-driven cancer research and is fuelled by the emergence of significant cancer-focused initiatives like The Cancer Genome Atlas (TCGA) and Clinical Proteomic Tumour Analysis Consortium (CPTAC) and is being employed in innovative approaches for cancer prognosis, prediction, and treatment [36, 56]. Multi-omics data integration is crucial for advancing precision medicine, clinical applications, and biomarker discovery. By combining various levels of biological information, such as DNA sequence, gene expression, and medical images, omics data enable the identification of biomarkers for early disease diagnosis, treatment response, and disease classification [15].

With the advancement of deep learning (DL) and the increased availability of multi-omics datasets, DL as an integrative multi-omics data approach is primarily being considered for identifying disease subtypes or classifying subgroups, identifying potential biomarkers for diagnostics and driver genes for diseases, and gaining insights into disease biology. In biomedical research, there is growing interest in DL applications for omics data analysis. Omics data analysis often encounters challenges such as low signal-to-noise ratios as well as datasets with many variables and relatively few samples or high analytical variance. DL techniques have already demonstrated superior performance compared to previous statistical and non-DL methods in terms of sensitivity, specificity, and efficiency. Furthermore, DL algorithms can analyze each data type separately and integrate different omics types or other sources of information, such as medical images and clinical health records. DL algorithms enable the implementation of various integration strategies by offering flexible and explicit design options for network architectures. This analysis of big data and integration is encouraging the adoption of personalized medicine approaches, facilitating early disease detection and classification, and enabling tailored therapies based on each patient's biochemical profiles [21, 26].

Deep learning models, such as multilayer perceptrons, convolutional neural networks, and generative adversarial networks, have greatly enhanced the integration of multi-omics data [15]. Such complex models offer unparalleled benefits to clinical environments; they significantly enhance medical imaging and diagnostics, enabling accurate disease detection and image segmentation. Moreover, Personalized medicine benefits from these models through the analysis of genetic and clinical data to tailor individualized treatment plans. Extracting insights from Electronic Health Records (EHRs) becomes more efficient with natural language processing capabilities, improving clinical decision-making. Additionally, deep learning accelerates drug discovery and development by predicting drug interactions and effectiveness, ultimately transforming healthcare practices [39, 58]. On the other hand, deep learning models are prone to overfitting due to their flexibility and may operate on loose assumptions. Additionally, those models consist of numerous layers interconnected by nonlinear relationships, making it challenging to fully understand their decision-making process and are often considered less accountable, functioning as black box models that lack transparency in their decision-making process. To address this limitation, supplementary post-hoc explainable artificial intelligence (xAI) models are necessary for thorough analysis. Despite their power, neural networks can be less intuitive to human observers, requiring additional efforts for interpretation and validation in multi-omics research. Understanding black box models is crucial for clinicians to trust AI predictions and ensure patient safety. It enables them to detect and correct AI errors, thus preventing potential misdiagnoses. The unexplainability of these models limits clinicians' ability to explain treatment recommendations. Clinicians have an ethical responsibility to ensure AI-recommended treatments align with patients' values and needs. Additionally, comprehending these models helps mitigate psychological and financial burdens on patients by providing clear information. In addition, clinicians' understanding of black box models is essential for developing guidelines and standards for the ethical and effective use of medical AI tools [19]. Therefore, to make such models applicable clinically, interpretability and explainability are essential, especially when questions about accountability arise in case of errors. xAI has emerged as a principal approach in the ML research community. In medical imaging, researchers are increasingly leveraging xAI to clarify algorithm results. A good explanation provides insight into how a neural network reaches its decision, making the decision understandable [26, 55, 57, 61].

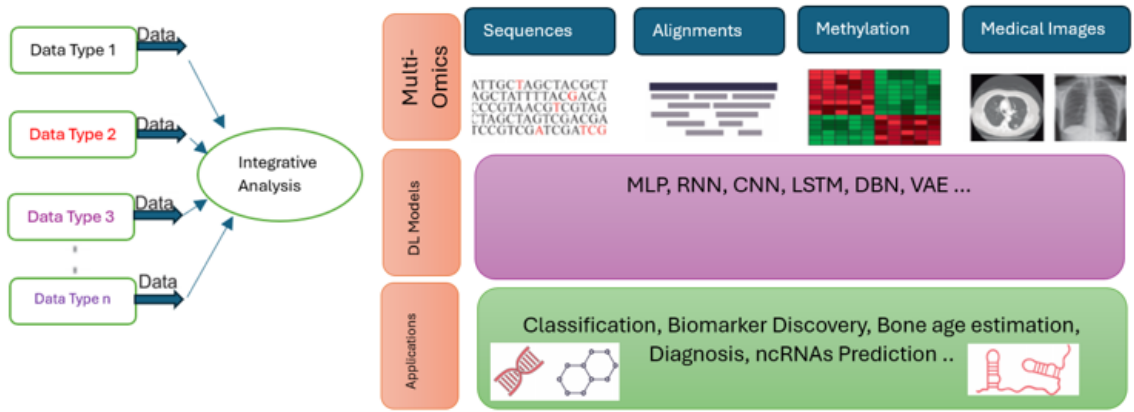

Fig. 2. Multi-omics data analysis represents a vertical integration approach where data from different layers are combined using various deep learning models. Some of the main applications achieved by such methodologies include classification, biomarker discovery, bone age estimation, diagnosis, and prediction of non-coding RNAs

## 2 Background

### 2.1 Overview on Multi-Omics Deep Learning Methods

Deep learning (DL) approaches leverage integrated Multiomics data to improve accuracy and efficiency in biomarker discovery and patient stratification [3, 15]. DL models have been successfully employed to classify cancer types based on gene expression patterns and methylation data, identify biomarkers associated with survival rates, and develop tools for cancer type classification using somatic mutations. Integrating multiple omics datasets allows for a more comprehensive understanding of disease mechanisms and facilitates the development of personalized treatment strategies tailored to individual patients' molecular backgrounds [15]. DL is a subset of machine learning that uses artificial neural networks (ANNs), which are inspired by biological neural networks. It has the ability to learn features across several layers. DL is increasingly recognized as a powerful approach for encoding and learning from heterogeneous and complex data, in both supervised and unsupervised settings. DL methods have made significant advancements in various artificial intelligence challenges and are increasingly being applied in biomedical research, particularly in omics data analysis. DL have demonstrated superior performance in terms of sensitivity, specificity, and efficiency compared to previous methods, particularly in overcoming challenges such as low signal-to-noise ratios and large datasets with a high number of variables. Moreover, with DL algorithms different omics layers and other sources of information, such as medical images or clinical health records can be integrated. This integration of big data analysis is driving the implementation of personalized medicine approaches, enabling early disease detection, classification, and personalized therapies tailored to each patient's biochemical background [27]. Figure 2 illustrates the Deep Learning Integration approach for multi-omics data, and Table 1 briefly describes common integrative deep learning approaches.

### 2.2 Overview of Explainable Artificial Intelligence Methods

The high dimensionality of omics data and the emergence of advanced AI, ML, and DL approaches has generated opaque black box models, prompting researchers to turn to explainable artificial intelligence (xAI) for analyzing omics data and uncovering insights into biological processes [56]. Explainability ensures the authenticity of model outputs by revealing the training data, fairness assessments, bias reduction efforts (Explainable data), the contribution of specific model features for specific outputs (Explainable predictions), and the composition of individual layers within the model and their contribution to the final output (Explainable algorithms) [36]. Researchers are compelled to opt for xAI methods due to several advantages. Firstly, xAI methods offer transparency in system processes, unlike conventional black-box AI models, by providing detailed insights into data preprocessing, model design, implementation, evaluation, and conclusions. This transparency allows for effective system configuration, improvement, versioning, and auditing. Secondly, xAI enhances explainability, instilling trust in users through detailed explanations of decision processes at every step, aiding in identifying and resolving model design issues. Thirdly, xAI models offer interpretability at each structural layer, enabling users to assess data quality, feature distribution, categorization, and classification, thereby fostering user confidence and trust in the system. Additionally, xAI models demonstrate high adaptability to emerging situations, facilitated by feedback techniques, allowing for modification and adoption of new features, especially crucial in the context of evolving medical scenarios [18, 44]. Overall, xAI methods ensure consistent model quality and applicability for long-term usage, addressing critical needs in various domains, including healthcare.

| DL Model | Description | Omics studies utilizing DL Models |
|----------|-------------|-----------------------------------|
| MLP (Multi-Layer Perceptron) | A type of feedforward artificial neural network with multiple layers of neurons. Each neuron in one layer is connected to every neuron in the next layer. | DGMP [50], a framework for identifying cancer driver genes by combining DGCN and MLP. |
| RNN (Recurrent Neural Network) | A neural network designed for sequential data, with connections forming directed cycles to maintain memory of previous inputs. | A study [49] investigated various RNN architectures focusing on gene expression data and evaluated them using multiple classification quality criteria. |
| LSTM (Long Short-Term Memory) | A type of RNN with memory cells to maintain information over long periods, addressing the vanishing gradient problem. | |
| CNN (Convolutional Neural Network) | Uses convolutional layers to learn spatial hierarchies of features from input data. | A multi-omics data integration study [1] utilized CNN for breast cancer stage prediction. |
| DBN (Deep Belief Network) | A generative model with multiple layers of hidden variables (latent variables) connected between layers but not within them. | A DBN-based model [43] was developed to detect Alzheimer's disease using DNA methylation and gene expression data. |
| VAE (Variational Autoencoder) | Learns a probability distribution over input data to generate new data samples from the learned distribution. | Used in various omics studies for dimensionality reduction [3, 31, 51]. |

Table 1. Deep Learning Models and their Applications in Omics Studies

As shown in Figure 3, explainability methods are categorized based on four main criteria: the scope of explainability, implementation, applicability, and explanation level [38]. The scope of explainability determines

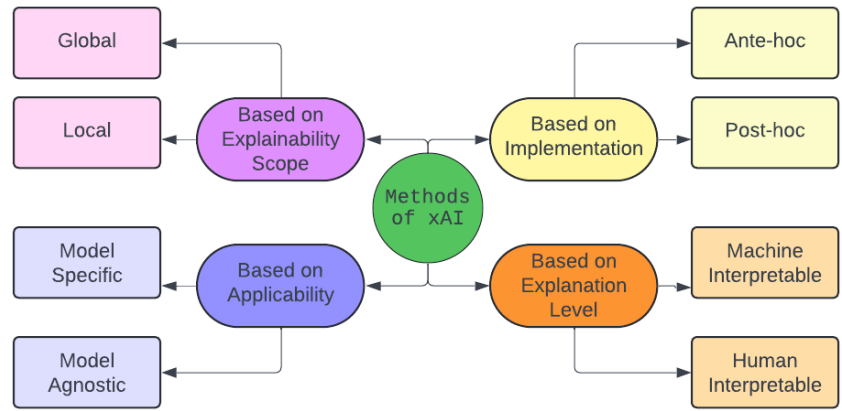

Fig. 3. xAI Classification Methods

whether an xAI method provides a global understanding of the entire model or a local explanation focused on specific input instances. Global explainability offers a comprehensive view of a model's overall behavior, while local explainability provides insights into how individual inputs affect the model's outputs [38, 52]. Regarding implementation, xAI methods are divided into ante-hoc and post-hoc approaches. Ante-hoc methods are integrated directly into the model design whereas post-hoc methods are employed after model training to interpret complex models that lack inherent explainability, using techniques such as Grad-CAM, LRP, LIME, and Saliency Maps to mimic and clarify the model's behavior post-creation [2, 38]

The applicability of xAI methods can be either model-specific or model-agnostic. Model-specific methods are tailored to particular model types, offering intrinsic explanations that align closely with the specific model's structure and functions. Conversely, model-agnostic methods like LIME, LRP, and SHAP can be applied across different models without modifying their performance, making them suitable for post-hoc explanations in a wide range of scenarios [8, 38]. Lastly, xAI methods are categorized based on the level of explanation they provide, which can be either machine-interpretable or human-interpretable. Machine-interpretable explanations are crafted for algorithmic analysis, focusing on technical details and model behavior, while human-interpretable explanations are designed to be easily understood by people, often using visualizations, natural language, or simplified concepts to make the model's decision-making process transparent and accessible [38]. This comprehensive categorization highlights the diverse approaches to xAI, each designed to meet specific needs for model interpretability and transparency.

## 3 Methodology

The review was conducted using a set of relevant keywords Figure(4a), such as "Multi-omics," "xAI," "Explainable," "Deep Learning," "Omics," "Genomics," "Cancer," and "Biomarkers. The keywords was searched across multiple academic databases, including Google Scholar, Springer Link, IEEE, and ScienceDirect. More than 600 papers were returned. The papers were screened by keywords and titles, narrowing them down to 300. These 300 papers were further screened based on their titles and abstracts. From these, 22 studies were selected based on inclusion criteria, which required the use of an omics or multi-omics framework combined with explainable AI (xAI) methods. The selected papers were then analyzed qualitatively to extract insights and conclusions pertinent to the study. The bar chart Figure 4b illustrates the number of records identified from various sources, including Google Scholar, Springer Link, and Science Direct. The word cloud (Figure 5) visually represents the most frequently used terms in the abstracts of the selected studies, emphasizing the central role of "model," "biomarker," "data," and "method." These terms suggest a strong focus on developing predictive models that integrate diverse biological data types to identify potential biomarkers. Words like "prediction," "cancer," and "feature" further indicate the emphasis on using these models for disease prediction and classification, particularly in cancer research. The word cloud also highlights the importance of data-driven approaches, machine learning techniques, and the integration of multi-omics data to advance personalized medicine and improve treatment outcomes. Additionally, selected papers employing xAI methods were categorized and analyzed as shown in Figure 6 allowing for a comprehensive examination of the techniques used in the field of xAI for multi-omics and providing valuable insights into their applications and effectiveness.

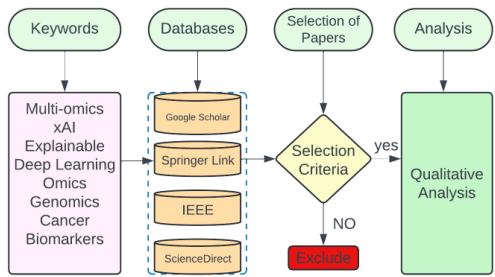

(a) Flowchart of the research process from keyword selection to qualitative analysis of selected papers.

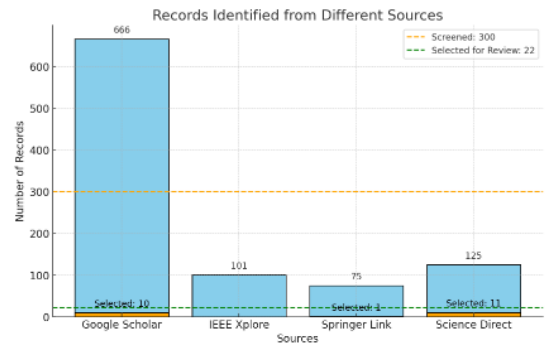

(b) Bar chart showing the number of records identified and selected for review from different databases.

Fig. 4. Research workflow (a) and a bar chart of selected studies (b).

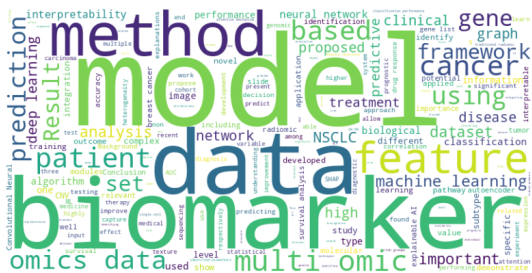

Fig. 5. Word cloud illustrating the key terms and concepts frequently associated with multi-omics data analysis .

## 4 Review of Explainable AI Methods for Multi-Omics

### 4.1 Model Agnostic Approaches

#### 4.1.1 SHAP Shapley additive explanations.
Shapley values are based on coalitional game theory, where the feature values of a data instance act as players in a coalition, and Shapley values tell us how to fairly distribute the prediction among the features [40]. SHAP explains the prediction of an instance $x$ by computing the contribution of each feature to the prediction. Shapley values provide explanations by assigning a value called weight to each feature for a particular prediction. SHAP considers all possible predictions, using all possible combinations of inputs, to guarantee consistency and local accuracy. KernelSHAP and TreeSHAP are two variants of SHAP; KernelSHAP is a model-agnostic approach based on LIME and Shapley values concepts, while Tree SHAP computes exact SHAP values for decision tree-based models [38]. The SHAP explanation is specified by the formula:

$$\phi_i = \sum_{s \subseteq d \setminus \{i\}} \frac{|s|! \, (|d| - |s| - 1)!}{|d|!} \left[ F_{s \cup \{i\}}(x_{s \cup \{i\}}) - F_s(x_s) \right].$$

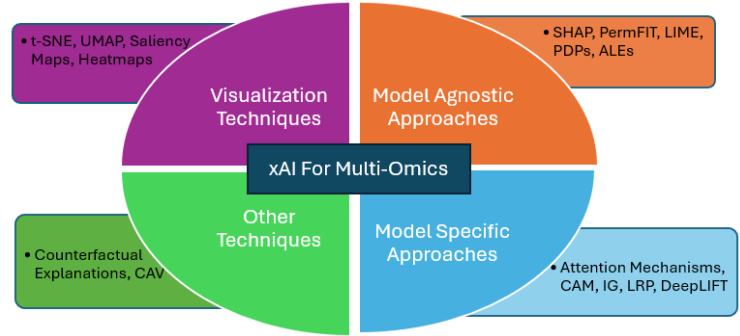

Fig. 6. xAI Methods for Multi-Omics

It represents the SHAP value for feature i, indicating its contribution to the prediction for the instance x . d is the set of all features in the dataset, and s is a subset of d excluding feature i. GradientSHAP is an xAI method implemented based on Shapley values. It approximates the Shapley value by computing an expectation of gradients over a distribution of baselines. This implementation extends the concept of IntegratedGradients from utilizing a single baseline to utilizing several baselines sampled over a distribution and computing an expectation of gradients for them. GradientSHAP is defined as:

$$GSi(x) = \int_{x'} \left[ (x_i - x_i') \int_{\alpha=0}^{1} \frac{\partial F(x' + \alpha(x - x'))}{\partial x_i} d\alpha \right] P_B(x') dx'$$

This formula represents the GradientSHAP value for feature i in instance x, x" is the baseline sample, is the input sample and F is the neural network model [30, 38, 40].

SHAP have been incorporated into several multi-omics investigation studies table 2, each with distinct approaches. In a study [30] aimed at uncovering important DNA methylation biomarkers for non-small cell lung cancer (NSCLC) classification, an xAI-guided deep learning network was developed. The framework consists of two blocks: Block A for NSCLC instance classification using a model that combines an autoencoder with a feed-forward neural network with methylation data as input, and Block B for biomarker discovery using various xAI methods including IntegratedGradients, GradientSHAP, and DeepLIFT to eventually identify a set of 7 significant biomarkers. The datasets used for experimentation were generated by the TCGA Research Network, specifically for NSCLC classes, and were extracted from LinkedOmics and cBioportal portals. The set of discovered biomarkers was evaluated for their classification performance, clinical efficacy, and association with biological pathways. The model's performance was assessed using 10-fold cross-validation to ensure the robustness and reliability of the results, with accuracy as the primary metric. Another study [31] introduced XL1R-Net; an Explainable AI-driven improved L1-regularized deep neural architecture for NSCLC biomarker identification. It is a novel approach designed to identify clinically relevant biomarkers for NSCLC, distinguishing between its major subtypes, LUAD and LUSC. The model utilizes gene expression data, focusing on copy number variations (CNVs) as input. The xAI methods used include IntegratedGradients, GradientSHAP, and DeepLIFT, to compute the contribution score of each gene. The model successfully identified twenty biomarkers, with nine overlapping with existing literature and seven novel NSCLC-relevant biomarkers discovered. The identified biomarkers show potential for targeted drug development and significant relevance in predicting patient survival. The model demonstrates superior classification performance for subtype classification of NSCLC instances compared to standard neural networks and achieves a state-of-the-art classification accuracy of 84.95% in subtyping NSCLC into its primary subtypes LUAD and LUSC.

xAI-CNV Marker is a proposed framework to discover robust biomarkers for breast cancer subtypes [51] , the framework identified CNV biomarkers linked to enriched pathways and prognostic capabilities and achieved a 5-fold cross-validation classification accuracy comparable to state-of-the-art methods. The framework utilizes deep learning for breast cancer classification, it consists of two main sub-modules: an autoencoder for data compression and a feed-forward classifier for cancer subtype classification. The xAI part of the framework incorporates a discovery algorithm that employs various explainable models including Gradient Input, Integrated Gradient, Epsilon Layerwise Relevance Propagation, DeepLIFT, and SHAP, for interpreting the model's decisions and identifying relevant biomarkers. Gene set analysis, survival analysis, and draggability analysis was applied to establish the clinical relevance of the identified biomarkers, in addition, the identified biomarkers was validated on an independent cohort (METABRIC dataset) to confirm their efficacy in distinguishing breast cancer subtypes . AutoGGN [34] is another framework that integrates omics data with molecular networks to enhance classification tasks. The framework underwent evaluation across three classification tasks: single-cell embryonic stage, pancancer type, and breast cancer subtyping, demonstrating superior performance compared to other methods. The framework utilized several datasets, including the TCGA Breast Cancer Subtype dataset, which comprised data from 396 patients with overlapping feature data and subtype information; the TCGA Pan-cancer dataset, consisting of 5780 patient samples with both gene expression and somatic mutation profiles across 24 cancer types; and a Mouse Single-cell Transcriptomes dataset, containing single-cell RNAseq profiles from 116,312 single cells across 10 developmental stages of mouse embryos, with a final dataset containing 10,000 single-cell samples. SHAP was employed for feature importance estimation, highlighting the crucial features for the classification tasks.

A study [3] aiming to stratify cancer patients into high-risk and low-risk groups using multi-omics integration utilized a multi-omics feature learning framework with three main components: feature extraction, tensor analysis, and risk prediction. Autoencoders were employed for dimensionality reduction, followed by fusing the processed data into a tensor using the CP decomposition method to facilitate the extraction of interpretable latent factors. The risk prediction aspect of the framework employed classification for tumor purity prediction in breast cancer datasets and clustering for stratifying cancer risk groups based on integrated multi-omics data, including SCNV, miRNA, RNAseq, and methylation. Clustering methods were evaluated using survival analysis, and SHAP was utilized to understand the contribution of biomarkers to the latent variables of omics data. Additionally, t-SNE was employed to visualize the latent variables extracted from the multi-omics data to aid in identifying clusters of samples with similar latent variable values. The study focused on Breast Invasive Carcinoma and Glioma datasets, as they contained more than 600 clinical samples, making them suitable for analysis. The proposed framework outperformed MOFA – A framework for unsupervised integration of multi-omics datasets - in terms of statistical significance and clustering accuracy.

The presented models encounter several limitations that affect their performance and generalizability. These include the small number of samples in some omics datasets as well as the complexity of biological systems which poses challenges in capturing the heterogeneity of all biological processes. Moreover, the limited biological interpretation in such studies which focuses primarily on computational aspects necessitates further research to interpret the biological significance. Issues related to demographic variations and potential biases in the datasets, particularly concerning the under representation of certain ethnic groups, may hinder the applicability of the model's findings across diverse populations. The "black-box" nature of deep learning models presents challenges in comprehensively understanding decision-making processes, despite efforts to incorporate explainable AI (xAI) tools. Dependency on high-quality data for training, concerns regarding data availability and privacy, as well as challenges associated with model complexity and interpretability, further constrain the models' utility and accessibility. While promising results are observed in specific classification tasks, the ability of the models to generalize across diverse biological datasets and conditions remains to be fully evaluated.

*4.1.2 LIME Local Interpretable Model-Agnostic Explanations.* LIME is an open-source framework aimed at illuminating the decision-making processes of machine learning models, thereby fostering trust in their utilization. Its "local" nature involves analyzing specific instances rather than providing a broad explanation for the model's behavior. It explains how a particular instance is categorized. Being "interpretable" implies that users should grasp the model's operations. For example, in image classification, it reveals which parts of an image influence predictions, while with tabular data, it highlights influential features. "Model-Agnostic" denotes its applicability to any black-box algorithm, past or future, disregarding whether the model's internal workings are transparent, it treats all models as black boxes. The term "explanations" refers to the output generated by the LIME framework. LIME is designed to interpret the decision-making processes of machine learning models across various data types and can effectively provide explanations for complex deep learning models by focusing on the neighbourhood of a specific instance [11, 40]. A framework for radiomics to predict Radiation Pneumonitis (RP) [53], a radiation-induced chest irritation, utilized radiographic images from 122 patients undergoing chest radiotherapy for various thoracic malignancies. Different Gradient Boosting Machines (GBMs) such as XGBoost, LightGBM, and CatBoost were used, and PyRadiomics, a Python package designed for analyzing radiographic images, was employed for feature extraction.

| Reference | Model Name / Purpose | Data / Data Source | Method / ML or DL Method | xAI Method |
|---|---|---|---|---|
| [30] | (NSCLC) classification Model / uncovering important RNA-Seq and CNV biomarkers for NSCLC | RNA-Seq, CNV and methylation data of TCGA NSCLC classes | Deep learning-based autoencoder and a feed-forward deep neural network | IntegratedGradients, GradientSHAP, and DeepLIFT |
| [31] | XLIR-Net / NSCLC Classification and biomarker identification | CNVs profiles of NSCLC patients generated by the TCGA Research Network | L1-regularized deep neural network | IntegratedGradients, GradientSHAP, and DeepLIFT |
| [51] | xAI-CNVMarker for breast cancer subtype Classification/ discover robust biomarkers for breast cancer subtypes | TCGA BRCA CNV Dataset | Autoencoder for data compression and a feed-forward classifier for cancer subtype classification | Gradient Input, Integrated Gradient, Epsilon Layerwise Relevance Propagation, DeepLIFT, SHAP |
| [34] | AutoGON a framework that integrates omics data with molecular networks / Enhancing classification tasks | TCGA Breast Cancer Subtype dataset, TCGA Pan-cancer dataset, Mouse Single-cell Transcriptomes dataset | Graph convolutional neural networks (GCNs) | SHAP |
| [3] | Identifying of cancer risk groups | SCNV, methylation, miRNA and RNAseq for Breast cancer and Glioma | Autoencoder / Tensor Analysis / classification/ clustering | SHAP, t-SNE plots |
| [59] | Drug response/ medical imaging classification/Identifying biomarkers for complex diseases | (RPPA) Data from three kidney cancer studies in (TCGA), HIITChip Atlas Microbiome Data employed for studying the human intestinal tract microbiome | DNN, RF, SVM | PermFIT, SHAP |
| [53] | Framework for RP Prediction | Radiographic images from 122 patients undergoing chest radiotherapy | XGBoost and DNN | LIME |
| [42] | QLattice, a symbolic-regression-based algorithm | Proteomics Alzheimer's disease data | Symbolic-regression-based algorithm | PDP |
| [14] | Study focusing on paediatric neuroblastoma | CT-based morphologic and radiomics features | XGBoost | ALE |

Table 2. Omics Studies Utilizing various xAI Methods in their Framework

| Reference | Model Name / Purpose | Data Source | ML / DL Method | xAI Method |
|---|---|---|---|---|
| [17] | MultiGATE/ Cancer Subtype identification method | TCGA Datasets of mRNA, DNA methylation and miRNA for various cancer types | Graph autoencoder network | Omics-level attention mechanism |
| [13] | MOGLAM / Classification Task / identifying important biomarkers | mRNA expression, DNA methylation, and miRNA expression data from KIPAN dataset for kidney cancer type classification | Dynamic graph convolutional network with feature selection (FSDGCN) | Multi-omics attention mechanism (MOAM) |
| [33] | MOMA / Interpretation and classification of multi-omics data/ predicting phenotypes and detecting modules containing genes related to phenotypes | Gene expression and DNA methylation from TCGA data sets for 34 classes (33 types of cancer and 1 normal) | 2 Module encoders (One for each omic) | Attention Layer/Assigns higher weights to relevant modules for disease prediction |
| [35] | MFAGN/ Survival and drug response prediction in digestive system tumors (DST)/ significant biomarkers detection. | Gene expression, DNA methylation, and copy number variation from TCGA and GDSC databases | View Correlation Discovery Network (VCDN) to combine omics data | Graph Attention Network (GAN) to extract features from omics data |
| [5] | Omics-CNN / High-throughput omics data classification/ Identify biosignatures for Ischemic Stroke (IS) and COVID-19 infection | One transcriptomics and one proteomics dataset alongside clinical and demographic data | Univariate and multivariate techniques for dimensionality reduction/ 1D CNN | Weighted Class Activation Mapping (GW-CAM) for determining important features |
| [4] | DeepsInsight-3D / Patient-specific anticancer drug responses | Multi-omics profiles such as gene expression, copy number alterations (CNA), and somatic mutations | Data-to-image conversion / CNN for automatic feature extraction analysis | Class Activation Mapping (CAM) for identifying gene activation regions |
| [6] | Head and neck cancer outcome prediction / Enhance the performance of traditional radiomics | Pre-treatment CT images from 106 patients | CNN | Gradient Class Activation Maps (CAM) / Highlighting important genes in images |

Hussein et al.

Table 2. Continued - Omics Studies Utilizing various xAI Methods in their Framework

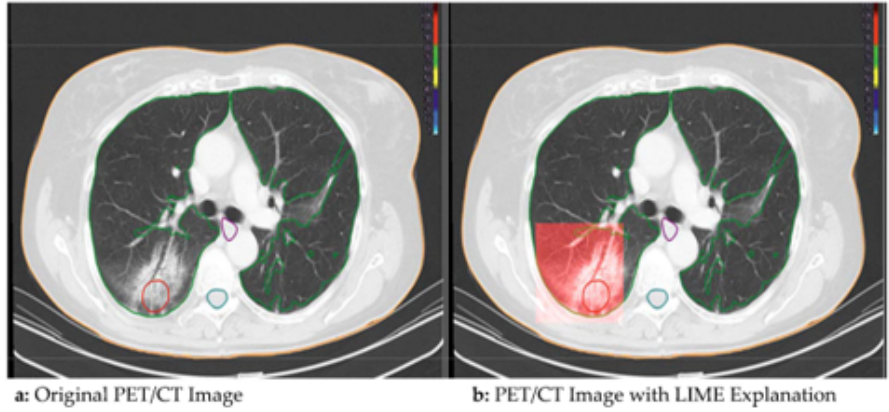

**a:** Original PET/CT Image    **b:** PET/CT Image with LIME Explanation

Fig. 7. Figure 7(a) shows a PET/CT image of a lung with radiation pneumonitis, while Figure 7(b) includes a LIME explanation highlighting the affected areas. A red frame at the image's lower portion emphasizes these regions. This visualization method helps healthcare professionals understand the model's decision-making process, validating findings and improving patient care by linking predictive accuracy with clinical understanding [53]

SHAP and LIME methodologies were used to enhance model interpretability, providing insights into feature importance and localized explanations for predictions. Figure 7 shows how LIME was utilized to show RP affected areas in a PET/CT lung image.

LIME has some notable limitations. On a local level, while LIME effectively calculates the influence and importance of features for individual predictions, it fails to connect prediction probabilities with feature probability graphs, leading to potential confusion as the feature importance scores do not sum up to the prediction probabilities. On a global level, LIME's utility is limited since it lacks built-in functionality for global evaluation, necessitating manual extraction and analysis of multiple observations to gain global insights. Additionally, the absence of comprehensive documentation for the tabular explainer can cause misinterpretations, as users often rely on third-party explanations from online articles and videos. Addressing these limitations would involve incorporating detailed guidance into LIME's documentation and developing performance indicators for global comparison [11].

*4.1.3 PDP and ALE.* A Partial Dependence Plot (PDP) illustrates the marginal effect that a feature set $S$ has on the predicted outcome of a machine learning model. It helps to identify whether the relationship between the target and a feature is linear, monotonic, or more complex. PDPs work by marginalizing the model's output over the distribution of features in set $C$, allowing the function to reflect the relationship between the features in set $S$ and the predicted outcome. This marginalization process results in a function dependent solely on the features in $S$, encompassing interactions with other features. The partial function reveals the average marginal effect on the prediction for given values of features in $S$. A key assumption of PDPs is that the features in $C$ are not correlated with those in $S$; if this assumption is violated, the calculated averages may include unlikely or impossible data points, thus distorting the PDP. The partial dependence function $\hat{f}_{x_S}(x_S)$ is estimated using the following formula:

$$\frac{1}{n}\sum_{i=1}^{n} f(x_S, x_{C_i}^{(i)}),$$

the partial dependence function shows the average predicted outcome when the features in $S$ are fixed at specific values $x_S$, while the other features (in set $C$) vary over their marginal distribution [40]. Accumulated Local Effects (ALE) plots describe how features influence the prediction of a machine learning model on average. ALE plots are a faster and unbiased alternative to Partial Dependence Plots (PDPs). They provide a robust method for interpreting the influences of predictors in machine learning models. ALE plots focus on calculating the average change in the model's prediction as a feature varies, while averaging out the effects of other features. This approach addresses the limitations of PDPs by not assuming independence between features, thus avoiding biased interpretations in the presence of correlated predictors. ALE plots visualize and clarify feature interactions and their impact on model predictions, especially in scenarios involving high-dimensional data spaces, like multi-omics [63]. In a study [42], QLattice, a symbolic-regression-based algorithm, was applied to Proteomics Alzheimer's disease data. The dataset was split, allocating 80% for training and 20% for testing. From the QLattice analysis on the training partition, the ten best unique models were derived. The best model was then selected based on the lowest Bayesian Information Criterion (BIC) score, indicating the best balance of model complexity and fit. This chosen model incorporates MAPT (tau protein), age at CSF collection, and LILRA2 as inputs, using addition operations to estimate the probability of Alzheimer's disease (AD) in patients. Notably, MAPT, a known AD biomarker, emerged as a consistent variable in the highest-scoring models from the QLattice, though other features varied among the models. A Partial Dependence Plot (PDP) was used to analyze the influence of MAPT on AD risk, showing how changes in MAPT levels alone could predict changes in AD risk, independent of interactions with other features. Another study [14] focusing on paediatric neuroblastoma, the MYCN gene amplification—known to indicate aggressive tumour behavior and poor survival—was examined using CT-based morphologic and radiomics features. Six morphologic and 107 quantitative gray-level texture radiomics features were extracted from volumes of interest manually outlined in the imaging data. The final predictive model utilized the eXtreme Gradient Boosting (XGBoost) algorithm. To interpret the influence of predictive features within this model, Accumulated Local Effects (ALE) plots were employed to provide insights into how each variable influenced the prediction of MYCN amplification status. ALE plots revealed that having more than six image-defined risk factors (IDRF) was associated with a higher probability of MYCN amplification, highlighting the utility of ALE plots in uncovering complex relationships within high-dimensional imaging data.

*4.1.4 Permutation Feature Importance.* PermFIT is a permutation-based feature importance method designed to identify significant features in machine learning models. It is a computationally efficient method as model refitting is not needed when it is employed. PermFIT calculates feature importance scores by assessing the expected increase in prediction errors when the values of a feature are permuted, thus indicating the feature's impact on the model's outcome. PermFIT [59] utilizes various visualization techniques such as bar plots and heat maps to output the calculated feature importance scores. PermFIT is characterized by enhanced statistical robustness and valid statistical inference as it does not require prior knowledge of feature distributions and incorporates permutation tests with cross-fitting. In a study [59] utilizing PermFIT for the identification of important biomarkers for complex diseases via machine learning models, datasets such as Reverse Phase Protein Arrays (RPPA) Data from three kidney cancer studies in The Cancer Genome Atlas (TCGA), HITChip Atlas Microbiome Data employed for studying the human intestinal tract microbiome, TCGA Kidney Cancer Data specifically analyzed for identifying important features associated with kidney cancer and BMI Levels from HITChip Atlas. PermFIT was implemented across (DNN, RF, SVM) models to estimate and test the feature importance. It outperformed existing feature selection methods, demonstrating its robustness and effectiveness. PermFIT's performance improvement is restricted by the inherent limitations of the machine learning models it is applied to, such as RF's inefficiency in modelling interaction terms, and its effectiveness in deciphering complex genetic architecture may be limited for traits with strong gene-gene interactions due to the underlying model's capabilities.

## 4.2 Model Specific Approaches

*4.2.1 Attention Mechanisms.* Attention mechanisms mimic our brain's way of focusing on specific things while ignoring others. In deep learning, they facilitate the processing of large amounts of information by focusing on the most important parts. Attention is becoming a key part of explaining how computers explain images, understand languages, and make decisions. By using attention, deep learning models can enhance their performance across various tasks by effectively managing the information to which they attend, thereby improving efficiency and accuracy. Attention mechanisms are usually categorized based on four criteria, as shown in table 3 [62].

| Attention Criterion | Types | Notes |
|---|---|---|
| Softness of Attention | Soft/Hard, Global/Local | **Soft Attention**: Assigns continuous weights to all parts of the input.<br>**Hard Attention**: Focuses on a subset of inputs.<br>**Global Attention**: Considers all input positions.<br>**Local Attention**: Focuses on a window of input positions. |
| Forms of Input | Item-wise, Location-wise | **Item-wise**: Each feature/item is treated individually.<br>**Location-wise**: Focus on the spatial or temporal location of features. |
| Input Representations | Distinctive, Self, Co-attention, Hierarchical | **Distinctive Attention**: Attends to different types of input separately.<br>**Self-Attention**: Computes attention within the same input sequence.<br>**Co-Attention**: Attends to two different input sequences simultaneously.<br>**Hierarchical Attention**: Multi-level attention mechanism that combines different granularities. |
| Output Representations | Single-output, Multi-head, Multi-dimensional | **Single-output Attention**: Produces a single context vector.<br>**Multi-head Attention**: Uses multiple attention heads to capture different aspects.<br>**Multi-dimensional Attention**: Produces multi-dimensional outputs for complex tasks. |

Table 3. The four attention mechanism criteria.

Attention mechanisms were utilized in various multi-omics studies table 2, MultiGATAE [17] is a novel cancer subtype identification method using multi-omics data and attention mechanisms. Eight TCGA Datasets were used for experiments, the study focused on various cancer types, including kidney renal clear cell carcinoma (KIRC), breast invasive carcinoma (BRCA), colon adenocarcinoma (COAD), skin cutaneous melanoma (SKCM), glioblastoma multiforme (GBM), lung squamous cell carcinoma (LUSC), liver hepatocellular carcinoma (LIHC), and ovarian serous cystadenocarcinoma (OV). The developed model employs a graph autoencoder network alongside K-means clustering for subtype identification. MultiGATAE constructs a similarity graph from multi-omics data, which is then processed by a graph autoencoder network consisting of a graph attention network and an omics-level attention mechanism to derive embedding representations. Subsequently, K-means clustering is applied to these embeddings to identify cancer subtypes. According to the model authors, MultiGATAE demonstrates superior performance in identifying distinct subtypes with varying survival outcomes across multiple cancer datasets and outperforms state-of-the-art methods in this task. MOGLAM [13] is another framework for disease classification task. The model takes in mRNA expression, DNA methylation, and miRNA expression data from KIPAN dataset for kidney cancer type classification, SCC dataset for pan-cancer classification and BRCA dataset for breast invasive carcinoma. It consists of three modules: dynamic graph convolutional network with feature selection (FSDGCN), multi-omics attention mechanism (MOAM), and omic integrated representation learning (OIRL). FSDGCN Learns optimal omic-specific embedding information while identifying important biomarkers. MOAM Adapts the attention mechanism to assess the importance of embedding information from different omics, enabling the model to emphasize more relevant omics data for downstream classification tasks. OIRL uses common and complementary information between different omics data types, aiming to enhance the model's classification performance by integrating multi-omics data in a more meaningful way. The model's performance was evaluated using accuracy (ACC), average F1 score weighted by support (F1weighted), and macro-averaged F1 score (F1macro), performance was compared against eight other methods, showing superior results in terms of ACC and F1 scores on the used datasets.

MOMA [33] is a multi-task attention learning algorithm designed for the interpretation and classification of multi-omics data. It focuses on identifying disease-related biological pathways through multi-omics module analysis and classification. The model comprises three stages: building a module for each omic dataset, highlighting important modules across omics data using module attention, and multi-task learning for each dataset. MOMA aims to provide interpretability alongside performance which is crucial in biology and medicine for accurate decision-making related to diagnosis and treatment. It can predict phenotypes and detect modules containing genes related to phenotypes by utilizing a module attention matrix to identify the most relevant modules for a specific phenotype. When compared with other models, it demonstrates robust performance and good generalization even in the presence of heavy noise . The Multi-omics Fusion Graph Attention Network (MFGAN) [35] integrates multiple omics data for survival and drug response prediction in digestive system tumors (DST). It employs the Graph Transformer (GT) for learning new graph structures and the Graph Attention Network (GAN) to extract features from omics data. Additionally, it incorporates the View Correlation Discovery Network (VCDN) to combine omics data features. The model uses omics data from TCGA and GDSC databases, including gene expression, DNA methylation, and copy number variation and predicts survival risk and drug response. Evaluation via 5-fold cross-validation, with a focus on mean AUC and AUPR scores, demonstrates improved performance compared to other methods.

The methods discussed exhibit several limitations inherent to the analysis of cancer omics data. These include challenges related to sample size, potential bias introduced by excessive clustering, and the lack of differentiation in node weights affecting feature relevance. Additionally, issues arise from unspecified omics types, potentially compromising the model's efficiency to consider correlations between omics data types. Furthermore, limitations such as the "curse of dimensionality," reliance on labelled data, and performance dependency on data quality and completeness are noted. While the models offer advancements, concerns persist regarding their complexity, accessibility, and interpretability, as well as variations in performance across different cancer types, feature selection oversights, and dataset scale constraints. Future efforts aim to address these limitations by incorporating additional omics data types and expanding dataset scales for improved generalization.

### 4.2.2 *Class Activation Mapping* .

Class Activation Map (CAM) [10, 38] is an explanation method suitable for convolutional neural networks with global average pooling. It enables CNNs to perform object localization, alongside classification, without the need for bounding box annotations. CAM highlights the discriminative region of the input image to identify its class, utilizing activation maps of the last convolutional layer to train linear classifiers for each class and estimate the final class. The important image regions for prediction are identified by projecting the weights of the output layer onto the convolutional feature map. CAM requires retraining linear classifiers for each class, making it time-consuming. Gradient-weighted CAM (Grad-CAM) is a proposed method aimed at reducing the time complexity of CAM. Grad-CAM is suitable for any convolutional neural network architecture and utilizes the gradients of the target feature flowing into the final convolutional layer to visualize the class activation maps. Additionally, it highlights the important regions in the input image of the selected feature. However, it's noteworthy that Grad-CAM may fail in localizing objects with multiple occurrences of the same class.

Omics-CNN [5] is a Convolutional Neural Network-based framework designed for high-throughput omics data classification. The study utilizes one transcriptomics and one proteomics dataset, along with clinical and demographic data, which include various characteristics such as age, BMI, pre-existing medical conditions, and laboratory measurements like C-reactive protein (CRP), absolute neutrophil count, and D-dimer. The objective is to develop diagnostic models and identify biosignatures for Ischemic Stroke (IS) and COVID-19 infection. The framework integrates dimensionality reduction, preprocessing, clustering, and explainability techniques. To tackle the high-dimensional nature of omics data, both univariate and multivariate techniques for dimensionality reduction are employed. The preprocessing pipeline involves normalization of datasets from different sources to generate a consistent expression matrix. Architecturally, the Framework comprises convolutional layers, pooling layers for input down-sampling, dense layers for learning complex feature relationships, and a softmax activation function for class prediction. Gradient Weighted Class Activation Mapping (GW-CAM) analysis is applied to identify the most important features for classification, aiding in the discovery of diagnostic biomarkers. Notably, the framework demonstrates high accuracy in identifying biosignatures for Ischemic Stroke and COVID-19, achieving accuracies of 96% and 95.41%, respectively. Another novel approach is DeepInsight-3D [4], designed to predict patient-specific anticancer drug responses from multi-omics data by employing data-to-image conversion for CNN application. The framework converts multi-layered tabular omics data into corresponding images, which are organized and coloured before being fed into a CNN for automatic feature extraction and analysis. DeepInsight-3D provides a methodological basis xAI, facilitating the identification of biologically relevant gene sets and the discovery of functional annotations important for classifications, aligning with known pathways and mechanisms in cancer and drug resistance. Utilizing patient-derived xenograft (PDX) datasets, including multi-omics profiles such as gene expression, copy number alterations (CNA), and somatic mutations, the model's performance was evaluated on seven datasets from The Cancer Genome Atlas (TCGA) and PDX, showcasing its capability in analyzing complex multi-omics data. Additionally, training and test datasets were constructed from Genomics of Drug Sensitivity in Cancer (GDSC) and TCGAPDX, resulting in more than seven datasets for comprehensive testing and validation of the model's predictive capabilities. The model architecture employs manifold techniques for data transformation alongside hull algorithms for mapping elements to pixel locations. It incorporates Class Activation Mapping (CAM) for identifying gene activation regions and an element decoder for isolating a subset of genes and then uses CNN for automatic feature extraction and analysis of the constructed images. DeepInsight-3D's performance was compared against several benchmark methods, including MOLI, NMF, feed-forward net, and others, using AUC as a key metric, demonstrating its robustness to high dimensionality and complex relationships between variables.

In a study [6] utilizing CNN and CAM for head and neck cancer outcome prediction, the aim was to detect image patterns that might not be covered by a traditional radiomic framework and to enhance the performance of traditional radiomics. A CNN framework was designed for medical image analysis, with the dataset consisting of pre-treatment CT images from 106 patients, identical to that used in a benchmark study. Gradient Class Activation Maps (CAM) were employed to illustrate the areas of the tumour images that were most crucial for the model's decision-making, highlighted in red on the images. These maps were then combined with the original CT images to aid researchers and clinicians in understanding how the model interpreted the images to make predictions, providing a visual explanation of the model's thought process. The CNN model outputs included predictions on distant metastasis, loco-regional failure, and overall survival, with AUC values of 0.88, 0.65, and 0.70, respectively, demonstrating its capability to predict cancer treatment outcomes based solely on CT images. In addition to CAM, feature activation maps are used as tools to trace back the spatial location of regions responsible for signature activation and to help in identifying the spatial-anatomical locations, enhancing interpretability in medical imaging. Activation maps have shown utility in differentiating histological subtypes of non-small cell lung cancer (NSCLC) by highlighting the importance of peritumoral regions in radiomics-based predictions [12].

The discussed studies face various limitations. The Omics-CNN framework, for instance, requires significant computational resources due to the large set of features in omics data. There is also a trade-off between model simplicity and performance when using multivariate dimensionality reduction methods. Furthermore, identifying meaningful biomarkers is challenging due to the complexity and noise inherent in omics data. Similarly, DeepInsight-3D faces constraints such as limited sample sizes, the complexity of anticancer drug response, and the need for sufficient training samples to achieve accurate estimations. These factors highlight potential challenges in practical applications. Furthermore, while Radiomics Feature Activation Maps offer interpretability benefits, they have several limitations. The model struggles to implement combination approaches for outcomes other than distant metastasis. Its reliance on a dataset of only 300 patients may limit generalizability. Training the model de novo without fully capturing the potential of transfer learning is another drawback. Moreover, the focus on pre-segmented tumor volumes simplifies the task but overlooks the variability in clinical segmentation.

### 4.2.3 *Integrated Gradients and Layerwise Relevance Propagation (LRP)*.

Integrated Gradients is an attribution method used in Explainable AI to determine the importance of individual features in a model's prediction. It calculates the integral of the gradients of the model's output with respect to the input features, taken along a straight path from a baseline (usually a zero vector or a neutral input) to the actual input. This method provides a way to attribute the change in the model's prediction to the change in the input features, ensuring that the attributions are both complete and consistent. In multi-omics analysis, Integrated Gradients can be particularly useful for understanding the deep learning models used for the integration process by revealing how different input features contribute to the final prediction [41]. Layerwise Relevance Propagation (LRP) is an Explainable AI method used to decompose the prediction of a neural network by attributing the relevance of the prediction back through the layers to the input features. By propagating the prediction score backwards through the network, LRP redistributes the relevance from the output to the input using specific propagation rules that ensure the conservation of the total relevance. This method helps in understanding which features contribute most to the prediction, offering insights into the decision-making process of complex Deep learning models [48].

A study [22] utilized Graph Convolutional Neural Network (GCNN) and Layer-wise Relevance Propagation (LRP) for stable feature selection and biomarker discovery in breast cancer. The study addressed instability in feature selection experienced with traditional methods. In the proposed framework, GCNN structured gene expression data using molecular network information, enhancing the model's ability to identify relevant features, while LRP explained model decisions by attributing relevance scores to input features, making gene prediction interpretation easier. Using a large breast cancer dataset, GCNN+LRP provided the most stable gene lists. The study also found that GCNN+SHAP is useful for impactful features in classification performance but less stable than LRP. GCNN+LRP demonstrated the most stable and interpretable feature sets, outperforming SHAP and traditional approaches, aiding in consistent biomarker identification across datasets and better understanding of breast cancer. Another Study , the GENIUS framework [37] integrates multi-omics data using advanced deep learning models tailored for image analysis. To enhance the detection of hidden, nonlinear patterns, this framework transforms multi-omics data into spatially connected images where each gene is represented as a pixel, allowing convolutional layers to analyze these spatial connections. The framework comprises an encoder, which compresses genomic information into a latent vector, a decoder, which reconstructs the image from the latent vector, and convolutional layers that extract information from the reconstructed image using a series of convolutional and max-pooling layers. Integrated Gradients (IG) are then used to assign attribution scores to each feature, allowing the interpretation of the model's predictions. The IG method evaluates the relative contribution of individual features, providing attribution scores for each gene in the transformed genome image. This helps identify which genes drive the model's predictions and are likely associated with disease progression. The study demonstrated significant improvements in predicting metastatic cancer progression from primary tumors. The framework was validated on multi-omics datasets from The Cancer Genome Atlas and two independent cohorts representing different stages of bladder carcinoma.

Regarding limitations, the GENIUS methodology may not fully satisfy the criteria of stability, interpretability, and classification performance simultaneously. Perturbed features can lead to artifacts, making it unclear if differences in feature impact originate from the data, model misbehaviour, or explanation method issues. In the other GCNN+LRP Framework there is a bias toward highly expressed genes in LRP-selected features, which may not correlate with impactful classification performance. The findings regarding GCNN+LRP's suitability for biomarker discovery are context-specific and may not generalize across different datasets or cancer types without further validation. The transformation of multi-omics datasets into images might introduce biases or lose information, and the framework's performance with different types of omics data needs extensive validation.

*4.2.4 DeepLIFT.* Deep Learning Important Features (DeepLIFT) is a widely used explanation framework for deep neural networks, providing insights into the importance of each feature in model predictions. It operates by comparing the model output to a reference output, based on differences in input from a reference input [9]. The reference input serves as a default or neutral input chosen to suit the problem at hand. DeepLIFT distinguishes between positive and negative contributions and assigns contribution scores to neurons based on the difference-from-reference quantity. Specifically, it computes the difference-from-reference quantity ($\Delta t$) as the difference between the target output and the reference output, then assigns contribution scores ($C_{\Delta n_j} \Delta t$) to the set of neurons necessary and sufficient to compute the target output, ensuring that their sum equals the difference-from-reference quantity. This methodology enables DeepLIFT to interpret the impact of individual features on model predictions in a transparent and interpretable manner.

DeepLIFT's approach allows information propagation even when the gradient is zero, a feature particularly advantageous in Recurrent Neural Networks (RNNs) where saturating activations like sigmoid or tanh are common, the method also mitigates the potential bias introduced by treating bias terms in the same way as gradients multiplied by inputs. Still, several questions remain open, such as how to effectively apply DeepLIFT to RNNs, how to empirically determine a suitable reference from the data, and how to optimize the propagation of importance through 'max' operations, as found in Maxout or Maxpooling neurons, beyond utilizing gradients alone.

In a study [23] conducted to demonstrate that integrating TCGA pan-cancer data improves survival analysis performance, meta-learning with Cox hazard loss was employed. DeepLIFT was utilized for model interpretability alongside methods like univariate Cox Lasso, iCluster, and PARADIGM for multi-omics data integration. Multi-omics cancer survival analysis outperformed the use of transcriptomics or clinical data alone. Additionally, employing DeepLIFT facilitated establishing a correlation between variable importance assignments and gene co-enrichment, indicating that genes with higher and similar contribution scores tend to be enriched together in the same sets. Meta-learning models exhibited superior performance over direct learning models in predicting cancer survival, as evidenced by higher C-index values across different cancer types when integrating multiple omics datasets.

## 4.3 Visualization Techniques

In the context of multi-omics research, saliency maps are powerful tools for interpreting deep learning models by highlighting the most influential features, such as specific genes, proteins, or metabolites, that drive the model's predictions. This method involves calculating the gradient of the model's output with respect to each input feature. This gradient-based approach allows for the creation of a visual map that indicates which parts of the input data are most critical for the model's decisions. By applying this method to multi-omics data, researchers can uncover key biomarkers and understand the complex interactions between different omics layers. Saliency maps thus enhance the transparency and interpretability of complex models, ensuring that predictions are based on biologically relevant information rather than artifacts or noise [29]. t-SNE (t-Distributed Stochastic Neighbor Embedding) and UMAP (Uniform Manifold Approximation and Projection) [32, 32] are two powerful techniques for dimensionality reduction and data visualization and can be utilized for high-dimensional biological data analysis. t-SNE is widely used to visualize complex data by converting high-dimensional data points into a 2D or 3D map, preserving local structures. It works by minimizing the Kullback-Leibler divergence between the probability distributions that represent pairwise similarities in the high-dimensional space and the lower-dimensional space. t-SNE however is computationally intensive and struggles with preserving global structures in large datasets. UMAP, addresses some of these limitations by preserving both local and global structures more effectively and efficiently. UMAP constructs a high-dimensional graph of the data and then optimizes a low-dimensional graph to be as structurally similar as possible. This makes UMAP particularly suitable for large-scale omics data integration, providing insightful visualizations that facilitate the discovery of patterns and relationships in complex biological systems. Heatmaps, as described by [16], are graphical representations that use colour to encode quantitative data, facilitating the visualization of data patterns and relationships. Originally developed for displaying gene expression data, heatmaps have evolved into a versatile tool for various fields. They represent data matrices with individual values encoded as colours, allowing for the easy identification of clusters, trends, and anomalies. The intensity of the colour typically represents the magnitude of the value, providing an immediate visual summary of complex data sets. This method of data representation is particularly useful for revealing hidden structures within the data, such as correlations between variables or distinct groupings of observations. The study by [47] represents a framework using deconvolution autoencoders to derive biological regulatory modules from single-cell RNA sequencing data. The autoencoder, trained with a Poisson loss function and a soft orthogonality constraint, processes raw expression data, bypassing the need for normalization. Saliency maps were used to evaluate the importance of each gene set on the hidden units of the autoencoder, aiding in the identification of key biological pathways that contribute to the data's representation. UMAP visualization was applied to both the original data and the autoencoder's representation layer, showing that the autoencoder effectively captures and preserves the structure of the data. Heatmaps displayed the impact of hallmark molecular pathways on the autoencoder's hidden units and highlighted the differences in pathway activities across various cell types and conditions, aiding in the identification of functional regulatory modules. This integrative approach enhances the understanding of cellular functions and regulatory mechanisms in single-cell data.

## 4.4 Counterfactual explanations and Concept Activation Vectors

Counterfactual explanations [25] focus on the minimal input feature changes needed to alter a model's prediction, providing insights into how a model's output would change if the given input were altered. Based on the idea of counterfactual thinking, which investigates alternatives to events that have occurred, these explanations offer a reasonable mechanism for understanding unknown phenomena by examining what would happen under different initial conditions. The purpose of counterfactual explanations is to enhance the interpretability of machine learning models by allowing users to understand the model's behavior through actionable changes. Beyond explainability, counterfactual examples can uncover hidden biases in training datasets and provide a multi-faceted analysis for diverse audiences. Effective counterfactual explanations are governed by multiple objectives, including connectedness, proximity, plausibility, stability, and robustness, to ensure a comprehensive understanding of the model's decision-making process.

CLARUS [24] is an Interactive Explainable AI Platform for Manual Counterfactuals in Graph Neural Networks and is designed with a frontend, backend, and an API to facilitate communication between these components. It employs Graph Neural Networks (GNNs) to analyze patient-specific Protein-Protein Interaction (PPI) networks, the user interface (UI) incorporates packages for graph visualization and API communication, while backend and API functionalities utilize libraries for GNNs and web services. CLARUS allows users to manually manipulate patient-specific PPI networks by posing counterfactual questions, such as adding or removing nodes and edges, to observe changes in GNN predictions and xAI relevance. It supports retraining the entire GNN model upon modifications in PPI network structures, facilitating a deeper understanding of model predictions through counterfactual analysis. This interactive exploration of counterfactual changes enhances users' causal understanding and trust in AI models by providing insights into hypothetical scenarios. Applied in the biomedical domain, CLARUS analyzes cancer patient data using multi-omics datasets from The Cancer Genome Atlas (TCGA), supporting binary classification tasks like distinguishing between Kidney renal clear cell carcinoma (KIRC) and multiple cancer types. This platform can be enhanced by adding and removing node and edge features, expanding the dataset collection, and enabling user-uploaded datasets for broader investigational use within CLARUS.

While counterfactual explanations can be valuable for interpretability, they may not generalize well across different models or datasets [25], thus limiting their applicability in diverse settings. The effectiveness of these methods in non-linear classifiers and complex DL models is often assumed rather than empirically validated, raising questions about their broad applicability. Additionally, the complexity of generating plausible and actionable counterfactuals can hinder their interpretability for non-expert users, potentially limiting their practical utility. Ensuring that counterfactual changes are understandable and meaningful to various knowledge groups remains challenging, as explanations must be tailored to different levels of expertise. Furthermore, counterfactual explanations might inadvertently reveal or amplify biases present in the training data, leading to ethical concerns regarding their use and interpretation.

Concept Activation Vectors (CAVs) [7] provide a bridge between human-interpretable concepts and the internal workings of deep learning models, allowing researchers to understand how specific concepts influence model predictions. CAVs are generated by training a linear classifier on the activations of a deep learning model using examples that represent a specific concept versus random examples. Creating a CAV involves three steps: first, selecting examples that represent the concept to be analyzed; next, choosing random examples to serve as a baseline; and finally, training a linear classifier to distinguish between the concept examples and the random examples. This classifier's decision boundary forms the CAV, representing the concept within the model's activation space. By projecting model activations onto these vectors, it is possible to measure the sensitivity of the model's output to the concept, thus quantifying the influence of the concept on the model's decisions. This method helps to make the model's decision-making process more transparent and understandable by linking high-level human concepts with low-level model features, providing a tool for debugging, bias detection, and model interpretation.

The DLS (Deep Learning Systems) framework for biomarker prediction [45] employs a two-stage deep learning system based on the Inception-v3 architecture. Stage 1 focuses on patch-level analysis of HE-stained slides to predict biomarker status (ER, PR, HER2), while Stage 2 aggregates these predictions for a slide-level summary using pathology report data. The framework utilizes paired HE and IHC slides with precise alignment for model training, and models were trained on over a billion patches from hundreds of slides. Interpretability techniques include Concept Activation Vector (CAV) analysis, saliency maps via SmoothGrad, and unsupervised clustering to explain features associated with biomarker predictions. The platform uses TCAV to quantitatively test model predictions' association with specific histologic concepts. Saliency maps highlight key regions impacting predictions, enhancing understanding at the pixel level.

Despite its usefulness, Concept Activation Vector (CAV) has several limitations. It relies on linear classifiers, which may not capture complex, non-linear relationships in high-dimensional data. The interpretability provided by CAV is limited to layers where activations are linearly separable, potentially overlooking deeper, nuanced interactions. Its effectiveness depends on the careful selection of concept patches, which requires curation and may introduce bias if not representative. Additionally, the method assumes that the model's internal representations align with human-interpretable concepts, which may not always be the case [45].

## 4.5 Limitations of Studies Utilizing xAI for Multi-Omics

The application of explainable AI (xAI) in multi-omics research is faced with several challenges and limitations that can impact the performance and generalizability of the models used. A primary concern is the often small sample size in omics datasets, which can hinder the ability of models to capture the complexity and heterogeneity inherent in biological systems. This limitation can lead to overfitting and reduce the generalizability of the findings across different biological contexts. Additionally, the complexity of biological systems poses a significant challenge in accurately representing the diversity of biological processes. This complexity is often underrepresented in models, leading to a limited understanding of the full spectrum of biological interactions. Moreover, many studies in this domain focus predominantly on computational aspects, with insufficient attention to biological interpretation. Further research is required to reveal the biological significance of model outputs. Another critical issue is the potential for demographic variations and biases in the datasets. The underrepresentation of certain ethnic groups can result in models that do not generalize well across diverse populations, limiting the applicability of the findings. The "black-box" nature of deep learning models exacerbates these issues, as it hinders the comprehensive understanding of decision-making processes, despite the integration of xAI tools aimed at enhancing interpretability.

The dependence on high-quality data for training models introduces further limitations. Data availability and privacy concerns can restrict access to the comprehensive datasets needed for robust model training. Moreover, the complexity of these models often leads to challenges in their interpretability and accessibility, making it difficult for researchers to fully understand and trust the model outputs. The ability of xAI models to generalize across diverse biological datasets and conditions remains uncertain, particularly given the dependency on data quality and the challenges associated with the "curse of dimensionality." These models often struggle with performance consistency across different types of biological data, which can hinder their practical applicability.

Another significant limitation of using xAI in multi-omics research is the difficulty in interpreting complex interactions across different omics data layers. Although xAI aims to make model decisions more understandable, it often fails to clearly explain the complex relationships between various types of biological data, such as genes, proteins, and metabolites. This challenge is made worse by the need to combine data from different sources, each with its own way of being measured and varying in quality. As a result, the explanations provided by xAI can be fragmented or incomplete, making it hard to draw meaningful biological conclusions. Additionally, because each type of omics data has unique characteristics, xAI models might oversimplify these complexities, leading to a less accurate understanding of the underlying biology.

Future work in this field must address these limitations by expanding dataset scales, incorporating diverse demographic data, and developing more transparent and interpretable models. This includes efforts to enhance the biological relevance of computational findings and ensure that models can effectively handle the complexities of biological data.

## 4.6 xAI in Multi-Omics: Opportunities and Future Directions

In complex deep learning models, Explainable Artificial Intelligence (xAI) offers unprecedented insights into biological systems, enhancing the interpretability and reliability of such models. However, there are several challenges associated with xAI in multi-omics integration. The inherent complexity of multi-omics data, which involves interactions across various biological levels, makes it difficult for xAI methods to provide clear and concise explanations. Additionally, scalability issues arise as multi-omics datasets grow in size and complexity, necessitating scalable xAI methods to handle this increased demand. Another challenge is ensuring that the explanations provided are meaningful and consistent across different omics layers, which can be particularly difficult due to the heterogeneous nature of the data. Moreover, developing robust evaluation metrics to assess the effectiveness of these explanations is still an ongoing research area. Furthermore, implementing xAI techniques can be computationally intensive, requiring significant resources, especially for large-scale multi-omics data. The explanations provided by xAI methods can also be hard to interpret for non-experts, necessitating domain expertise, and making them less accessible to researchers without a strong background in both AI and biology. Additionally, when dealing with sensitive biological data, it is crucial for xAI methods to respect ethical standards and privacy concerns. Lastly, for an xAI-driven model to be accepted by the scientific community, it must produce reproducible and consistent results across different datasets and conditions.

The opportunities with xAI are numerous and promising for the field of multi-omics. One of the key advantages is enhanced interpretability, as xAI methods can make the decision-making process of machine learning models transparent. This transparency enables researchers to understand the contribution of different omics features, thereby facilitating more informed decisions. Furthermore, improved model trustworthiness is another significant benefit; by providing insights into model behavior, xAI can enhance a model's clinical applicability, helping to validate models and build trust among clinicians and patients. Additionally, xAI techniques can aid in feature reduction by identifying the most relevant features, which reduces dimensionality and improves model performance. Moreover, xAI can assist in error diagnosis by highlighting why a model makes certain errors, thus guiding data correction and model refinement. Finally, xAI can play a pivotal role in personalized medicine by interpreting how different omics features influence outcomes, thereby aiding in the development of personalized treatment strategies. Collectively, these opportunities demonstrate the profound impact xAI can have on advancing multi-omics research and applications.

Explainable Artificial Intelligence (xAI) presents a transformative approach to addressing the challenges of multi-omics integration. By enhancing interpretability, reducing dimensionality, and improving model reliability, xAI can accelerate the discovery of meaningful biological insights and the development of personalized medicine. Future research should focus on refining xAI techniques and exploring their applications across diverse multi-omics datasets.

## 4.7 Conclusion

Explainable Artificial Intelligence (xAI) presents both significant challenges and promising opportunities in the field of multi-omics. The primary challenge lies in the complexity and opaqueness of deep learning models, which are often regarded as black boxes due to their intricate layers and nonlinear relationships. This lack of transparency makes it difficult to understand the decision-making process, leading to potential issues with accountability and trust, especially in clinical applications. To address these limitations, xAI methods are essential for providing post-hoc explanations that make the outputs of these models more interpretable. Opportunities arise from the ability of xAI to enhance the interpretability of deep learning models, thus facilitating their integration into clinical practice. By clarifying how models derive their conclusions, xAI can help in identifying disease subtypes, discovering biomarkers, and understanding disease mechanisms, ultimately supporting personalized medicine. The integration of xAI in multi-omics not only improves model transparency but also fosters greater trust and adoption in biomedical research, paving the way for more accurate and individualized therapeutic strategies.

## Acknowledgments

This research is supported by an Australian Government Research Training Program Scholarship